\documentclass[preprint,5p,times,twocolumn]{elsarticle}

\usepackage{amssymb}
\usepackage[colorlinks]{hyperref}
\usepackage[ruled,linesnumbered,english]{algorithm2e}
\SetAlFnt{\small}
\SetAlCapFnt{\small}
\SetAlCapNameFnt{\small}
\SetAlCapHSkip{0pt}
\usepackage{color}
\usepackage[table]{xcolor}
\definecolor{dark-red}{gray}{0.60} 
\definecolor{medium-red}{gray}{0.80} 
\definecolor{medium-green}{gray}{1} 
\definecolor{dark-green}{gray}{1} 

\usepackage{array}
\newcolumntype{x}[1]{>{\centering\arraybackslash}p{#1}}
\usepackage{booktabs}
\usepackage{multirow}
\usepackage{subcaption}
\usepackage[flushleft]{threeparttable} 
\usepackage[justification=justified,singlelinecheck=false]{caption}

\usepackage{pgfplots}
\pgfplotsset{compat=1.14}
\usepackage[framemethod=TikZ]{mdframed}
\mdfdefinestyle{MyFrame}{%
    linecolor=black,
    innertopmargin=7pt,
    innerbottommargin=7pt,
    innerrightmargin=7pt,
    innerleftmargin=7pt
}

\newcommand{\RQone}{To what extent do the chapters have defined themes?}
\newcommand{\RQtwo}{To what extent are recipes related to the theme of the chapter where they were included?}
\newcommand{\RQthree}{To what extent are recipes suitable for inclusion in cookbooks?}
\newcommand{\RQfour}{To what extent do recipes have self-contained information?}
\newcommand{\RQfive}{To what extent do recipes meet all the properties \reciperelatednesstochapter, \recipeadequacy~and~\recipeselfcontainment?}
\newcommand{\RQsix}{What role can crowd cookbooks play in API learning?}
\newcommand{\RQseven}{What are the strengths, weaknesses and improvements that can be made to crowd cookbooks?}

\newenvironment{RQenvironment}[2]
{\vspace{10pt}\noindent{\small\textbf{RQ \##1: }}\textit{#2} \\ \begin{em}\normalfont}
{\end{em}}
\newcommand{\RQcitation}[1]{{\small\textbf{RQ \##1}}}

\newcommand{\chaptersemantics}{\textit{chapter semantics}}
\newcommand{\reciperelatednesstochapter}{\textit{recipe relatedness to chapter}}
\newcommand{\recipeadequacy}{\textit{recipe adequacy}}
\newcommand{\recipeselfcontainment}{\textit{recipe self-containment}}

\newenvironment{surveyquestionenvironment}
{\vspace{5pt}\noindent\textit{Survey question for participants.}\begin{em}\normalfont}
{\end{em}}

\newenvironment{responsetypeenvironment}
{\vspace{5pt}\noindent\textit{Response type.}\begin{em}\normalfont}
{\end{em}}

\newenvironment{contextenvironment}
{\vspace{5pt}\noindent\textit{Context.}\begin{em}\normalfont}
{\end{em}}

\newenvironment{findingsenvironment}
{\vspace{5pt}\noindent\textit{Findings.}\begin{em}\normalfont}
{\end{em}}

\newenvironment{statisticsenvironment}
{\vspace{5pt}\noindent\textit{Statistics.}\begin{em}\normalfont}
{\end{em}}

\newenvironment{participantscommentsenvironment}
{\vspace{5pt}\noindent\textit{Participants' comments.}\begin{em}\normalfont}
{\end{em}}

\newenvironment{additionalanalysisenvironment}
{\vspace{5pt}\noindent\textit{Additional analysis.}\begin{em}\normalfont}
{\end{em}}

\newcommand{\fivepointLikertscale}{(1) strongly disagree, (2) disagree, (3) neutral, (4) agree, and (5) strongly agree}

\usepackage{enumitem}
\setlist[1]{itemsep=-1pt}

\journal{Information and Software Technology}

\begin{document}

\begin{frontmatter}

\title{Bootstrapping Cookbooks for APIs from Crowd Knowledge on Stack Overflow}

\author{Lucas B. L. Souza}
\ead{lucas.facom.ufu@gmail.com}

\author{Eduardo C. Campos}
\ead{eccampos@ufu.br}

\author{Fernanda Madeiral}
\ead{fernanda.madeiral@ufu.br}

\author{Kl\'{e}risson Paix\~{a}o}
\ead{klerisson@ufu.br}

\author{Adriano M. Rocha}
\ead{adriano.comp2@gmail.com}

\author{Marcelo de Almeida Maia \corref{corauthor}}
\ead{marcelo.maia@ufu.br}
\cortext[corauthor]{Corresponding author.}

\address{Federal University of Uberl\^andia, Faculty of Computing, Uberl\^andia, Minas Gerais, Brazil}

\begin{abstract}
\textbf{Context:} Well established libraries typically have API documentation. However, they frequently lack examples and explanations, possibly making difficult their effective reuse. Stack Overflow is a question-and-answer website oriented to issues related to software development. Despite the increasing adoption of Stack Overflow, the information related to a particular topic (e.g., an API) is spread across the website. Thus, Stack Overflow still lacks organization of the \textit{crowd knowledge} available on it.

\textbf{Objective:} Our target goal is to address the problem of the poor quality documentation for APIs by providing an alternative artifact to document them based on the crowd knowledge available on Stack Overflow, called \textit{crowd cookbook}. A \textit{cookbook} is a recipe-oriented book, and we refer to our cookbook as \textit{crowd cookbook} since it contains content generated by a crowd. The cookbooks are meant to be used through an exploration process, i.e. browsing.

\textbf{Method:} In this paper, we present a semi-automatic approach that organizes the crowd knowledge available on Stack Overflow to build cookbooks for APIs. We have generated cookbooks for three APIs widely used by the software development community: SWT, LINQ and QT. We have also defined desired properties that crowd cookbooks must meet, and we conducted an evaluation of the cookbooks against these properties with human subjects.

\textbf{Results:} The results showed that the cookbooks built using our approach, in general, meet those properties. As a highlight, most of the recipes were considered appropriate to be in the cookbooks and have self-contained information.

\textbf{Conclusions:} We concluded that our approach is capable to produce adequate cookbooks automatically, which can be as useful as manually produced cookbooks. This opens an opportunity for API designers to enrich existent cookbooks with the different points of view from the crowd, or even to generate initial versions of new cookbooks.
\end{abstract}

\begin{keyword}
API documentation \sep Crowd knowledge \sep Stack Overflow \sep Cookbook
\end{keyword}

\end{frontmatter}

\section{Introduction}

Developers typically reuse predefined components through their {\it Application Programming Interfaces} (APIs). According to Parnas in the mid-90s, ``Reuse is something that is far easier to say than to do. Doing it requires both good design and very good documentation''~\cite[p.~224]{Brooks1995}.
Still nowadays, we have obstacles to reuse software, and many of these obstacles are somehow related to API documentation~\cite{Robillard2009,Robillard2011}.
Fortunately, other kind of documentation is emerging: an accessible structure of social media, such as wikis, blogs, forums, and question-and-answer (Q\&A) websites~\cite{Parnin2012} is 
redefining how developers learn, preserve and share knowledge on software development~\cite{Storey2017}. Barzilay et al.~\cite{Barzilay2013} even argue that this kind of knowledge has the potential to become a substitute to the official software documentation.

An important example of social media is Stack Overflow, which is a Q\&A website that allows the exchange of knowledge between developers.  
Related to API documentation, Parnin et al.~\cite{Parnin2012} verified that Stack Overflow provides high coverage of classes belonging to the APIs addressed in their study: for instance, 87\% of the Android API classes are referenced in questions on Stack Overflow. Delfim et al.~\cite{Delfim2016} extended the analysis performed by Parnin et al.~\cite{Parnin2012}, and  found that 69\% of the Android API top-level elements are referenced in questions of the type \textit{how-to-do-it} on Stack Overflow~\cite{Nasehi2012}, which is the question type where the asker provides a scenario and asks how to implement it. Interestingly, this type of question is adherent to the purpose of documenting how to use API elements.
Moreover, Treude and Aniche \cite{Treude2018} have shown that API documentation is dispersed among many sources, and still Stack Overflow plays a prominent role on it.

Those findings that Stack Overflow has a large volume of information on APIs, and  that conventional API documentation, when it exists, usually lacks examples and explanations, have motivated the usage of Stack Overflow content to produce API documentation. For instance, Treude and Robillard \cite{Treude2016} presented an automatic approach to augment API documentation with insight sentences on API elements mined from Stack Overflow, i.e., an approach to improve existing documentation. Moreover, Stack Overflow started a beta project \textit{Stack Overflow Documentation} that ran from July 2016 to August 2017, which was a repository of organized user-written code examples, supported by some explanation. Although the Stack Overflow organization has considered it as a good idea, they had to interrupt it, because of the high cost for content maintenance, among other reasons~\cite{SOdocumentation}.

In this paper, grounded on similar motivation, our goal is to provide a highly automated approach to select and organize the  content available in Stack Overflow, which is indeed a challenge because the content related to a particular API is spread across posts on Stack Overflow, lacking hierarchical organization. For instance, there are more than 41,000 discussion threads on Stack Overflow related to the Swing library organized with social tagging~\cite{Halpin2007}. To reach our goal, we developed a semi-automated approach that uses Stack Overflow content to construct a special type of documentation for APIs. We characterize this documentation as a \textit{cookbook}, because it is organized into chapters, and each chapter should bring together a collection of recipes about the same topic. Each recipe has a scenario and instructions on how to solve such scenario, containing explanations in natural language and source code snippets. Since we use the \textit{crowd knowledge} available on Stack Overflow to build cookbooks, we refer to them as \textit{crowd cookbooks}.

Cookbooks have already been widely used as an instrument to guide developers \cite{Martelli2002, Darwin2001, Stephens2006, Chang2012, Laurent2010, Polukhin2013, Wolff2011, Gundecha2012, Kuc2013, Sadun2013}. Note that cookbooks are meant to be used through an exploration process (browsing) instead of a search-driven one. For instance, in a searching process, a developer would be able to write a query, and a system would bring from Stack Overflow the best results for such query. Such searching process goes in the direction of \textit{automated on-demand developer documentation} \cite{Robillard2017} with recommendation systems, and a system of this type using Stack Overflow content is presented in \cite{Campos2016}. On the other hand, a \textit{crowd cookbook} could, for instance, be used by someone who wants to know what are the main themes of an API, which is specially useful for newcomers. According to Olston and Chi~\cite{Olston2003}, searching and browsing are complementary strategies, each one having advantages and disadvantages.  So, our goal is neither claiming  that crowd cookbooks are substitutes for searching-driven systems, nor claiming that the notion of cookbooks is better than other forms of documentation. Our goal is to demonstrate that adequate crowd cookbooks can be generated by our proposed automated approach.

This study extends our previous work~\cite{Souza2014b},  presenting the approach in full detail and providing a robust empirical evaluation with human subjects to produce evidence that the generated cookbooks are suitable for being bootstrapped towards a final version of an API documentation.

The main contributions of this work are:

\begin{itemize}
\item An approach to automate the construction of preprint cookbooks for APIs using the crowd knowledge available on Stack Overflow;

\item The definition of desirable properties that the generated cookbooks must meet, which are meant to measure the effectiveness of the proposed approach in organizing and filtering content. Assessing the technical content of the recipes (which is content mined from Stack Overflow) is not part of the scope of this work;

\item An in-depth empirical evaluation of the produced crowd cookbooks for widely used APIs\footnote{The cookbooks generated for the API subjects used in the evaluation are available at \url{http://lascam.facom.ufu.br:8080/QAWeb_sf/}.} conducted with a survey applied to graduate students and developers, considering the set of desirable cookbook properties.
\end{itemize}

The rest of this paper is organized as follows. 
Section~\ref{sec:background} setups a context for technical cookbooks and reviews  on technical documentation evaluation to contextualize the desired properties to assess our approach to generate cookbooks.
Section~\ref{sec:cookbook-construction} presents the proposed approach to build cookbooks.
Section~\ref{sec:evaluation-methodology} presents the methodology  to evaluate the cookbooks.
Section~\ref{sec:results-and-discussion} presents and discusses the results, followed by limitations and threats to validity that are presented in Section~\ref{sec:threats}.
The related works are discussed in Section~\ref{sec:related-work}, and the conclusions are presented in Section~\ref{sec:conclusions}.

\section{Background}\label{sec:background}

\subsection{Cookbooks}

Technical cookbooks are collections of recipes. As in food cookbooks, recipes are organized into topics/themes as the reader typically tries to find a recipe from its table of contents, which is a collection of chapters, each one for a topic/theme. In food cookbooks, a recipe is composed by ingredients and directions. A person without cooking knowledge would try to follow a recipe, and there is an increased chance of having success at preparing the food. The same idea is related to software: a developer, new to some programming task, would find a recipe that matches her task and would have an increased chance to complete it whether the directions are followed.

\begin{table*}[ht]
  \caption{Commercial versions of cookbooks.}
  \label{tab:cookbook-sample}
  \centering
  \small
  \begin{tabular}{l c c c c}
    \toprule
    Cookbook & \# Chapters & \# Recipes & \# Pages & Recipe structure \\
    \midrule
    Python Cookbook~\cite{Martelli2002} & 17 & 245 & 608 & \\
    Java Cookbook~\cite{Darwin2001} & 26 & 299 & 850 & Title + \\
    C++ Cookbook~\cite{Stephens2006} & 15 & 171 & 594 & Problem + Solution + \\
    R Graphics Cookbook~\cite{Chang2012} & 15 & 156 & 416 & Discussion + See also$^*$ \\
    jQuery Cookbook~\cite{Laurent2010} & 18 & 182 & 480 & \\
    \midrule
    Boost C++ Application Development Cookbook~\cite{Polukhin2013} & 12 & 95 & 348 & Title + \\
    Open GL 4.0 Shading Language Cookbook~\cite{Wolff2011} & 9 & 64 & 340 & Context + Getting ready$^*$ + \\
    Selenium Testings Tools Cookbook~\cite{Gundecha2012} & 11 & 91 & 326 & How to do it + How it works + \\
    Apache Solr 4 Cookbook~\cite{Kuc2013} & 7 & 106 & 328 & There's more$^*$ + See also$^*$ \\
    \midrule
    The Core iOS 6 Developer's Cookbook~\cite{Sadun2013} & 13 & 106 & 576 & Title \\
    \bottomrule
  \end{tabular}
\end{table*}

To characterize the organization of a recipe in the context of software development, we selected a sample of ten cookbooks related to different programming subjects (see Table \ref{tab:cookbook-sample}). Except for \textit{The Core iOS 6 Developer's Cookbook}, which contains recipes without internal structure, the other nine cookbooks follow a similar organization regarding the internal structure of recipes.
Table \ref{tab:cookbook-sample} shows the topics used to organize the content within recipes in column ``Recipe structure''. The ones that are marked with ``$^*$'' are not used in all recipes, but only when it makes sense---e.g. ``See also'' references other recipes in the same cookbook or external sources. The remaining ones, despite the names being different between the two separated groups of cookbooks, are close to be corresponding---``Problem''$\leftrightarrow$``Context'', ``Solution''$\leftrightarrow$``How to do it'', and ``Discussion''$\leftrightarrow$``How it works''. For this reason, we conclude that a typical recipe has, at least, the following contents:

\begin{enumerate}
\item \textit{Title}, which should be sufficiently succinct to appear adequately in a table of contents and sufficiently precise so the developer can match the title against the intended task to be performed. For instance, a recipe describing how to read from a file in Python has as title: \textit{Reading from a file}~\cite{Martelli2002}.

\item \textit{Problem description}, which also should be sufficiently succinct and relatively general so it can be matched to different contexts. For the description of the problem concerning reading a file, the problem description is \textit{You want to read text or data from a file}~\cite{Martelli2002}.

\item \textit{Solution}, which shows how to solve the problem. There may be different styles for the solution description: more succinct with a short statement reporting on what should be done~\cite{Darwin2001}, or a little bit more detailed with a simple example that ideally could be reproduced by the developer~\cite{Martelli2002}.

\item \textit{Discussion}, which may explain different facets for applying the solution, or may complement the solution with alternatives of implementation. The idea is that the solution should be kept as simple and general as possible, and more details could be presented in the \textit{Discussion} section.
\end{enumerate}

The structure of typical recipes directly matches question-and-answer pairs on Stack Overflow: the asker poses her question with a \textit{title} and describes the \textit{problem} that motivates the question, and responders provide \textit{solutions} for such problem.

\subsection{Cookbook assessment}

Although there is no framework to assess the quality of technical cookbooks, at the best of our knowledge, there has been some work on assessing technical documents. Arthur and Stevens~\cite{Arthur1992} developed a taxonomy for evaluating documentation, where an adequate documentation has four qualities: accuracy, completeness, usability, and expandability. Smart~\cite{Smart2002} studied quality factors for technical documentation in the context of software quality documents~\cite{Smart2002}, and proposed three dimensions to evaluate documentation quality: easy to use (i.e. task-oriented), easy to understand (i.e. concrete including examples), and easy to find (i.e. organized in a way that makes sense to the user). Later, Robillard~\cite{Robillard2009} observed that the following properties of software documentation are the most important: content (information in the document), organization (index, sections, subsections), use of examples, and being up-to-date.

To define our approach for cookbook generation, we took into account those properties. We rely on Stack Overflow to build the cookbooks and Robillard~\cite{Robillard2009} properties are, at least, partially met. The \textit{content} of the cookbooks is already assessed by the Stack Overflow community via the voting mechanism (which is used in our approach). Stack Overflow has content that is rich in examples and we only select answers containing source code snippets to compose the cookbooks to improve the easiness to understand how to solve a problem, thus the \textit{use of examples} property is also met. Since Stack Overflow allows the editing of posts, the property of \textit{being up-to-date} can be met. The \textit{organization} property is not always satisfied for the recipes, because not all answers in the cookbook are organized in sections, subsections. However, the cookbook itself partially meets this property since its recipes are grouped into chapters (themes).

Our approach considers, by definition, quality factors exposed in the literature, taking advantage of the nature of cookbooks and the content on Stack Overflow. Moreover, we organized the assessment of the cookbooks generated by our approach considering those quality properties (see our research questions in Section~\ref{sec:evaluation-methodology}). Recall that \textit{completeness} is out of the scope of cookbooks as they typically do not target completeness: their main purpose is to empower adopters of a language or API with the directions on how to solve programming tasks. Hence, we evaluate whether chapters are well defined, whether recipes are related to the chapter they are part of, whether recipes are suitable to be part of a cookbook, and whether recipes are self-contained.

\section{The Proposed Approach for Cookbook Construction}\label{sec:cookbook-construction}

\begin{figure}[!ht]
  \centering
  \begin{subfigure}[t]{\columnwidth}
  	\centering
  \includegraphics[scale=.55]{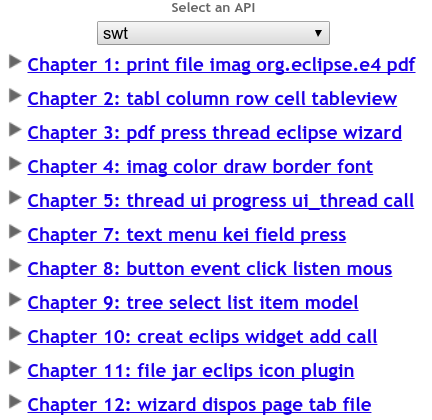}
    \caption{The SWT API Cookbook.}
    \label{fig:cookbook}
  \end{subfigure}
  \qquad
  \begin{subfigure}[t]{\columnwidth}
  	\vspace*{12pt}
    \centering
  \includegraphics[scale=.75]{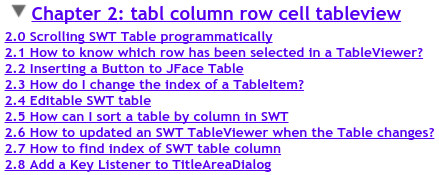}
    \caption{Chapter 2 from the SWT API Cookbook: SWT table component.}
    \label{fig:chapter2}
  \end{subfigure}
  \qquad
  \begin{subfigure}[t]{\columnwidth}
  	\vspace*{12pt}
 	\centering
  \includegraphics[scale=.4]{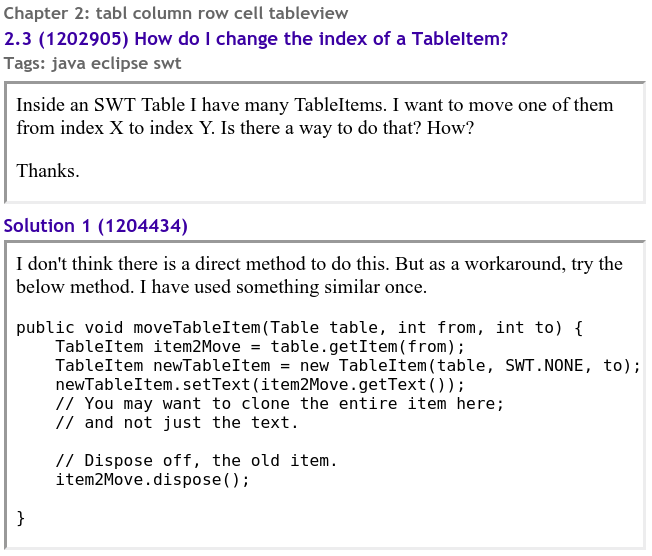}
  \caption{Recipe 2.3 from SWT API Cookbook.}
    \label{fig:recipe}
  \end{subfigure}
  \caption{A cookbook example: (a) cookbook summary, (b) recipes within a chapter, (c) a recipe.}\label{fig:cookbook-example}
\end{figure}

In this section, we present our approach to generate cookbooks from the knowledge available in Stack Overflow. A cookbook is a collection of chapters, and each chapter is a collection of recipes related to each other. In our context, a cookbook is related to a given API, and its chapters should be groups of related question-and-answer (Q\&A) pairs from Stack Overflow. A pair Q\&A  matches the concept of technical recipes, i.e., a pair problem/solution. An example of a cookbook for the SWT API, generated with our approach, is shown in Figure~\ref{fig:cookbook-example}. The chapters of the cookbook are presented in Figure~\ref{fig:cookbook}; the content of a chapter is shown in Figure~\ref{fig:chapter2}, containing nine recipes; the content of a recipe is shown in Figure~\ref{fig:recipe}, which provides a Q\&A pair from Stack Overflow.

\begin{figure}[ht]
  \centering
  \includegraphics[scale=0.52]{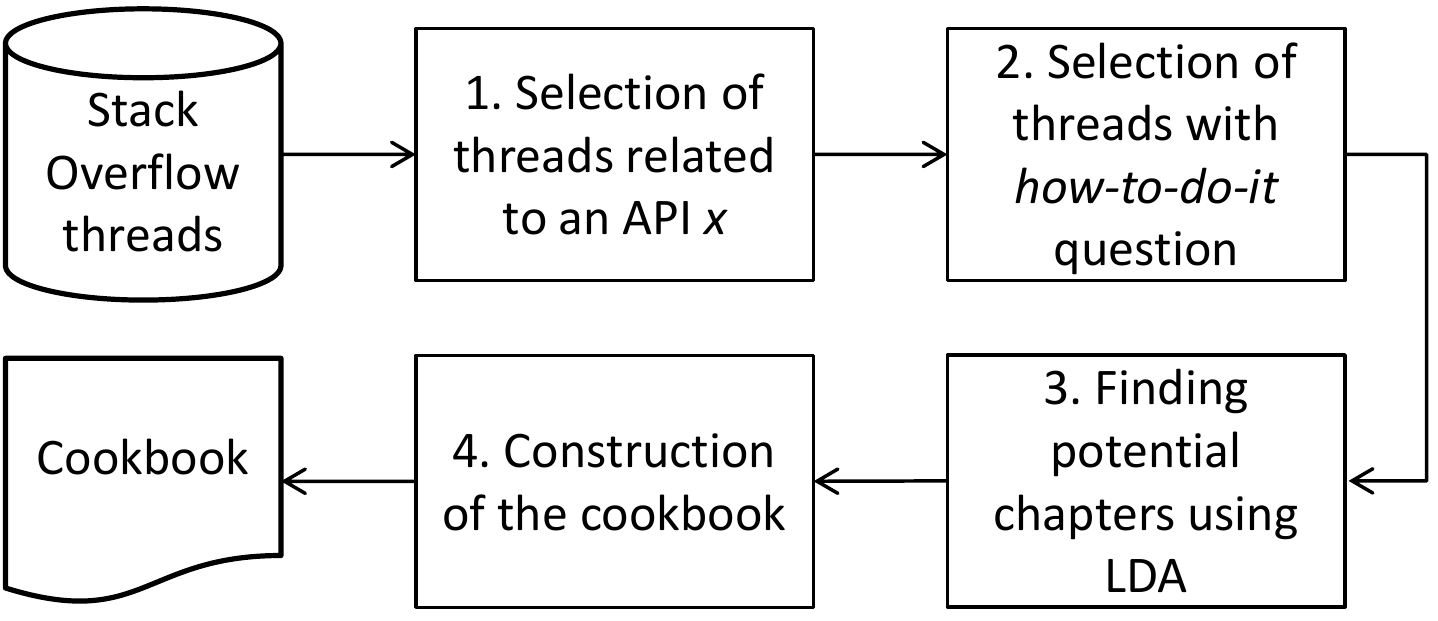}
  \caption{Overview of our cookbook construction approach.}
  \label{fig:overview-of-the-cookbook-construction-approach}
\end{figure}

Figure~\ref{fig:overview-of-the-cookbook-construction-approach} presents the overview of the main steps of our cookbook construction approach for a given API $x$. We obtained Stack Overflow threads (questions plus their respective answers) that are used to construct cookbooks from a dump\footnote{http://blog.stackoverflow.com/category/cc-wiki-dump/} that contains the entire content available on Stack Overflow since its creation in 2008 until September 2013. Then, given the obtained Stack Overflow threads and an API $x$, the cookbook construction approach consists of four main steps:

\begin{enumerate}
\item Selection of threads related to the API $x$ (Section \ref{sec:selection-threads-related-to-API});

\item Selection, among the previously selected threads, of threads having a particular type of question, \textit{how-to-do-it}, as this is the type of question suitable for cookbooks (Section \ref{sec:selection-how-to-questions});

\item Finding potential chapters for API $x$ using the topic modeling technique Latent Dirichlet Allocation (LDA)~\cite{Blei2003}, where each topic mined by LDA is a potential chapter to group related recipes in the cookbook being built (Section \ref{sec:lda});

\item Construction of the cookbook using the information generated by LDA, by determining its chapters and recipes (Section \ref{sec:algorithm-cookbook-generation}).
\end{enumerate}

\subsection{Selection of threads related to a given API x}\label{sec:selection-threads-related-to-API}

On Stack Overflow, the topic of a question, which shows the main technology or construct that such question revolves around (e.g., API), usually can be identified from the question tags~\cite{Nasehi2012}. Therefore, to select threads related to a given API $x$, we search for threads where their questions contain a tag related to the API $x$. For instance, to select threads on SWT API,  only threads with questions having the tag ``swt'' are selected.

\subsection{Selection of threads with how-to-do-it question}\label{sec:selection-how-to-questions}

Nasehi et al.~\cite{Nasehi2012} identified four question categories on Stack Overflow based on the main concerns of the askers and what they want to solve: \textit{debug-corrective}, \textit{need-to-know}, \textit{how-to-do-it} and \textit{seeking-different-solution}. They defined the \textit{how-to-do-it} category as the one where the asker provides a scenario and asks on how to implement it, which is very close to the idea of recipes from cookbooks. Thus, we are interested only in threads containing \textit{how-to-do-it} questions.

To verify the feasibility of constructing automatic classifiers for these categories of question, we performed experimentation with different classification mechanisms~\cite{Campos2014}. Based on the analysis of the most important features in that work, we constructed for this work a rule-based binary classifier to automatically filter only \textit{how-to-do-it} questions from the others. A question is classified as \textit{how-to-do-it} whether the three following conditions are satisfied:

\begin{enumerate}
\item It has the term ``how'' in its title or body;

\item It does not have in its body the presence of terms generally related to the \textit{debug-corrective} category: ``fail'', ``problem'', ``error'', ``wrong'', ``fix'', ``bug'', ``issue'', ``solve'', ``trouble'';

\item It does not have the word ``error'' in its code snippets (if any is present).
\end{enumerate}

The rules described above were defined after manually analyzing a random sample of 70 questions from Stack Overflow. We intended to differentiate between \textit{how-to-do-it} and \textit{debug-corrective} categories since we observed that the word ``how'' is also used in the latter category but with the intention to ask help on how to fix a problem.

To measure the accuracy of our rule-based classifier, we evaluated it by using a dataset we constructed in a previous work~\cite{Delfim2016}, which contains known \textit{how-to-do-it} and non-\textit{how-to-do-it} questions. Table~\ref{tab:classifier-evaluation-results} presents the size of such dataset, which contains 366 questions related to Android and 372 questions related to Swing API, totaling 738 questions, where 292 are \textit{how-to-do-it} ones. Table~\ref{tab:classifier-evaluation-results} also presents the results on overall accuracy, precision and recall, for both APIs. Note that our classifier has an overall accuracy of 77.91\% on the 738 questions, and the precision at classifying \textit{how-to-do-it} questions is 0.78.

\begin{table}[h!]
  \caption{Evaluation results on the performance of our rule-based classifier.}
  \label{tab:classifier-evaluation-results}
  \centering
  \small
  \begin{tabular}{p{.045\textwidth} x{.063\textwidth} x{.063\textwidth} x{.055\textwidth} x{.055\textwidth} x{.055\textwidth}}
    \toprule
    \multirow{3}{*}{API} & \multicolumn{2}{c}{Dataset} & \multicolumn{3}{c}{Evaluation results} \\
    \cmidrule{2-6}
    {} & \# Questions & \# How-to-do-it & \% Accuracy & \multirow{2}{*}{Precision} & \multirow{2}{*}{Recall} \\
    \midrule
    Android & 366 & 129 & 80.05 & 0.74 & 0.66 \\
    Swing & 372 & 163 & 75.81 & 0.81 & 0.58 \\
    \midrule
    Total & 738 & 292 & 77.91 & 0.78 & 0.62 \\
    \bottomrule
  \end{tabular}
\end{table}

\subsection{Finding potential chapters using Latent Dirichlet Allocation}\label{sec:lda}

A cookbook is organized into chapters, each one corresponding to a \textit{topic} of a given API. To identify the cookbook chapters, we use Latent Dirichlet Allocation (LDA), a topic modeling technique that automatically finds general \textit{topics} from a \textit{corpus of documents}, without the need of tags, supervised training, or predefined taxonomies~\cite{Blei2003}.
In our context, the corpus of documents is a set of threads, and the topics generated by LDA are the potential chapters for the cookbook. Note that we use LDA to group related Stack Overflow threads into topics.

To create the corpus of documents, for each previously thread identified as having \textit{how-to-do-it} question, we created a document containing the textual content of the thread by concatenating the question title, the body of the question, and the body of all answers.
Then, we preprocessed the corpus of documents similarly as Barua et al.~\cite{Barua2014}. 
For each document, we discarded code snippets, if any, because source code keywords (e.g., ``while'' loop) introduce noise into the analysis phase: all code snippets contain similar programming language syntax and keywords, which do not help the topic modeling technique to find meaningful topics~\cite{Barua2014,Thomas2011}. We removed HTML tags (e.g., \textless br\textgreater), because our focus is to analyze natural language (English) content. We also removed common English words (stop words), and applied the Porter stemming algorithm~\cite{Porter1997} to map the words to their base form. 
We used HTML Cleaner\footnote{http://htmlcleaner.sourceforge.net} to discard code snippets and to remove HTML tags, and Apache Lucene\footnote{http://lucene.apache.org/core} to remove stop words and stemming.

After the creation and preprocessing of the corpus of documents, we used the MALLET LDA implementation~\cite{McCallum2002}. Each topic mined by LDA possibly originates a chapter in the cookbook depending on the number of recipes included on it.
LDA characterizes every topic \textit{i} as a probability distribution over the terms.
Table \ref{table:frequenciaTermos} shows the probability distribution of the 10-top terms for a mined topic from the SWT API. The numbers indicate the relative importance: for instance, the term ``tabl'' is about ten times more important (considering the frequency) than the term ``checkbox''. Since we aim at assessing the raw content generated by our approach, we do not manually define a chapter name. Instead, we consider the top-5 terms of the topic as the chapter ``title'' in the rest of the paper.

\begin{table}[ht]
  \caption{Top-10 terms and their frequencies in the SWT API.}
  \label{table:frequenciaTermos}
  \centering
  \small
  \begin{tabular}{l c}
    \toprule
    Term & Relative importance \\
    \midrule
    tabl & 0.054351167 \\
    column & 0.025278521 \\
    row & 0.016221274 \\
    cell & 0.013537645 \\
    tableview & 0.011860377 \\
    select & 0.008617659 \\
    viewer & 0.007387662 \\
    jface & 0.006716755 \\
    item & 0.005710394 \\
    checkbox & 0.005263123 \\
    \bottomrule
  \end{tabular}
\end{table}

Additionally, as each document may refer to different topics, LDA also defines for each document \textit{j} a probability distribution over the topics, i.e., the adherence of the document to the topics.
Given a document, the sum of its adherences to the topics is always 1, thus the topic that has the highest adherence by such document is called its Dominant Topic. 

The number of topics ($K$) is a user-specified parameter and controls the granularity of  discovered  topics. Small values generate more generic topics and higher values generate more detailed topics. There is no single value of $K$ that is suitable for all situations and all datasets~\cite{Wallach2009}. 
In this study, we aim at  producing cookbooks with similar size as  commercial versions of cookbooks. So instead of trying to find an optimal value for $K$ as Panichella et al~\cite{Panichella2013}, we selected a sample of ten cookbooks, related to several programming subjects and counted the number of chapters of each one. This information is shown in Table \ref{tab:cookbook-sample}. The average number of chapters in our sample is 14.3, so we chose the value 15 for $K$, even though that value is not meant to be optimal.

\subsection{Construction of the cookbook}\label{sec:algorithm-cookbook-generation}

\begin{figure}[ht]
  \centering
  \footnotesize
  \begin{tikzpicture}
    \begin{axis}
    [ybar,
    width=1.05\linewidth,
    height=8cm,
    bar width=15pt,
    xlabel={Size ranges},
    xlabel near ticks,
    enlarge x limits=0.08,
    symbolic x coords={
    0 $\vdash$100,
    100 $\vdash$300,
    300 $\vdash$500,
    500 $\vdash$700,
    700 $\vdash$900,
    900 $\vdash$1\,100,
    1\,100 $\vdash$1\,300,
    1\,300 $\vdash$26\,900
    },
    xticklabel style={anchor=north east, rotate=30},
    ylabel={Percentage of questions (\%)},
    ylabel near ticks,
    ymin=0,
    ymax=25,
    xtick pos=left,
    ytick pos=left
    ]
      \addplot[draw=black, fill=white] coordinates {
      	(0 $\vdash$100, 1.55)
        (100 $\vdash$300, 17.76)
        (300 $\vdash$500, 21.35)
        (500 $\vdash$700, 16.44)
        (700 $\vdash$900, 11.76)
        (900 $\vdash$1\,100, 7.98)
        (1\,100 $\vdash$1\,300, 5.69)
        (1\,300 $\vdash$26\,900, 17.46)
      };
    \end{axis}
  \end{tikzpicture}
  \caption{Percentage of questions by size (number of characters) range.}
  \label{fig:percentage-of-questions-by-size-range}
\end{figure}
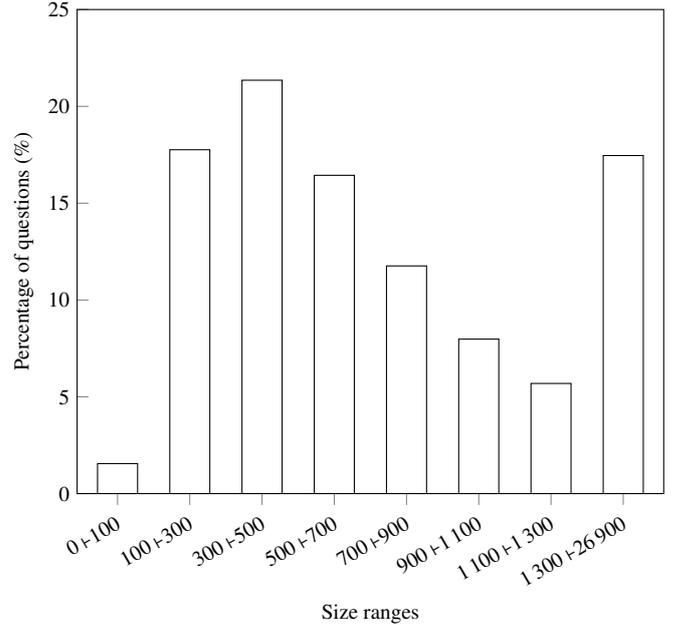

The generation of a cookbook for a given API requires LDA to be applied on the filtered corpus of documents (threads) whose questions are \textit{how-to-do-it}, resulting in \textit{K} topics (potential chapters), which should be filled with recipes. Consider that a recipe is a question-and-answer pair from the corpus. The reason to consider a recipe as a pair, rather than the entire thread, is due to the fact that a thread may have good and poor quality answers. Thus, when considering separated pairs, only eligible answers with high score are included in the cookbooks, as the score of a post is a proxy for its quality~\cite{Dalip2013}. We have defined some conditions that make a recipe (i.e., a Q\&A pair) eligible to be included in a cookbook:

\begin{itemize}
\item Condition \#1: The answer must have source code snippets, since programming by example is an intuitive way to learn both for novices and experts~\cite{Lahtinen2005}. Also, commercial cookbooks usually make code snippets publicly available on source code repositories\footnote{http://github.com/dabeaz/python-cookbook}. We identify the presence of snippets through the use of the HTML tags ``\textless pre\textgreater \textless code\textgreater...'';

\item Condition \#2: Both answer and question in a pair must not have dead links in its content. Prior work~\cite{Dellavalle2003} shown that dead links pose a major challenge to enhance educational resources as one cannot make use of a content that is no longer accessible. We used HttpUnit\footnote{http://httpunit.sourceforge.net/} to verify whether a link is dead;

\item Condition \#3: The question must not be too long. Verbose questions usually contain many inquiries or demonstrate the asker has difficulties to explain the problem. Even on Stack Overflow there is a policy for scope of questions. If the question is too broad, often verbose, users can put it ``on hold'' and eventually ``close'' the question. In other words, if the question is not problem specific, it cannot be answered on Stack Overflow. Further, 
Figure~\ref{fig:percentage-of-questions-by-size-range} shows a histogram with the percentage of questions in different size (number of characters) ranges. The data considered in this graph are all questions for SWT, STL and LINQ. The questions with size less than 1,300 characters comprise 81.4\% of the questions. Therefore we decided not to include pairs whose questions have size above 1,300 characters.
\end{itemize}

Since each individual post on Stack Overflow has its own score, we needed to define a metric that indicates the quality of the Q\&A pair as a whole. We consider that the answer tends to be more important than its question since the answer usually carries more information about the solution of the problem. Hence, we took into consideration the score of a pair as the weighted mean value between the individual scores of its question and answer. In our prior work~\cite[Section~4.3]{Campos2016}, we experimentally test several weights, and we found the optimal values as 0.3 and 0.7 for the respective weights. Different weight choices would lead to different results, but the above choice resulted in a balanced way to give more importance to answers, while still considering the relative importance of questions.

Algorithm~\ref{alg:cookbook-generation} is the pseudocode for building a cookbook $C$ for an API $x$. The goal is to build a cookbook that has at least $R$ recipes. To determine the value for this threshold, we used the information on the number of recipes in the sample of ten commercial cookbooks shown in Table \ref{tab:cookbook-sample}. As the smallest cookbook in this sample has 64 recipes, we decided to choose this value for the threshold $R$ to optimize recipe quality, while still guaranteeing that generated cookbooks contain at least as many recipes as conventional cookbooks.

\begin{algorithm}[ht]
	$maxRankPosAllowed \leftarrow getMaxRankPosAllowed(x)$\;
	$numRecipes \leftarrow 0$\;
	\While{$numRecipes < R$}{
    	$C \leftarrow new~Cookbook()$\;
 		$numRecipes \leftarrow 0$\;
		\ForEach{$doc \in How\-To\-Do\-It\-Corpus$}{
    		$domTopic \leftarrow getDominantTopic(doc)$\;
    		$adherence \leftarrow getAdherence(doc, domTopic)$\;
   			\If{$adherence \geq Ta$}{
                $pair \leftarrow getPair(doc)$\;
    			\If{$pair <> null$}{
    				$rankingPosition \leftarrow getRankingPosition(pair)$\;
                	\If{$rankingPosition \leq maxRankPosAllowed$}{
   						$includePairIntoC(C, pair, domTopic)$\;
   						$numRecipes \leftarrow numRecipes + 1$\;
   					}
                }
   			}
 		}
  		$removeSmallChapters(C)$\;
 		$maxRankPosAllowed \leftarrow maxRankPosAllowed + 10;$
	}
	\caption{Building a cookbook \textit{C} for an API \textit{x}}
    \label{alg:cookbook-generation}
\end{algorithm}

Although the minimum number of recipes is established by the threshold $R$, we could, in principle, include more recipes in the cookbook depending on their quality. Thus, for a given API, we build a ranking of Q\&A pairs based on their score, where the pair in position one in the ranking has the higher score. Then we establish the threshold \textit{maxRankPosAllowed} that defines the maximum position that a pair should have on the ranking, which prevents the selection of pairs under that quality threshold to be included in the cookbook.

\begin{figure}[ht]
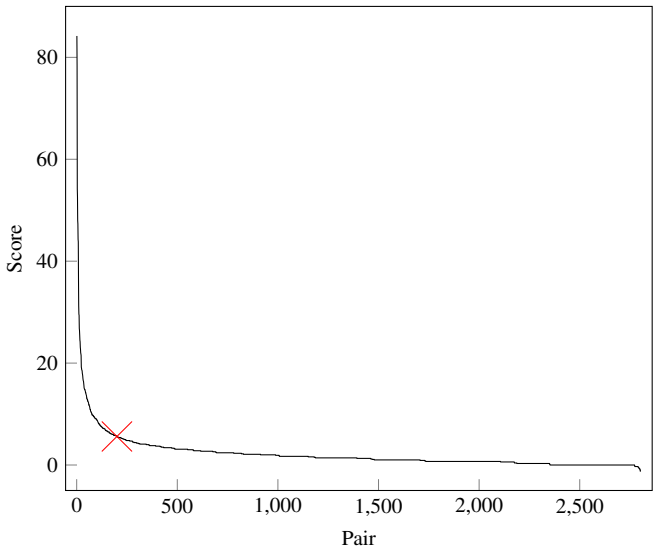

  \centering
  \footnotesize

  \caption{Score ranking of the Q\&A pairs for QT API.}
  \label{fig:score-ranking-QT}
\end{figure}

The initial value for this threshold (which is retrieved in line 1, Algorithm~\ref{alg:cookbook-generation}) is defined by analyzing the shape of the ranking curve. For example, Figure \ref{fig:score-ranking-QT} shows this curve for QT API. The idea of this analysis is to choose a cutoff point on the x-axis considering only the \textit{best-of-breed} pairs and discarding the long tail with lower score. For instance, for QT API, we choose the value 200. This analysis should be done for each API, because the number of threads can vary widely between APIs. We characterize the whole approach as semi-automatic only because of such cutoff point choice that has to be done manually.

\section{Evaluation Methodology}\label{sec:evaluation-methodology}

In this section, we present the methodology used to evaluate crowd cookbooks generated by using our approach presented in the previous section.

\subsection{Research questions}\label{sec:Research-questions}

To evaluate our crowd cookbook construction approach, we first formulated seven research questions. We further provide the answers for these research questions (in section \ref{sec:results-and-discussion}) based on a survey conducted with human subjects (participants), who analyzed sampled chapters and recipes and provided answers to our survey questions.

Five of the research questions are related to four desirable properties concerning crowd cookbooks (\RQcitation{1} to \RQcitation{5}), which are answered based on ratings given by participants in the study according to a 5-point Likert scale. The two additional research questions were posed to understand qualitative aspects on the characteristics of crowd cookbooks (\RQcitation{6} and \RQcitation{7}), which are answered based on free natural language texts given by participants in the study. We present each research question as follows, together with the associated survey question made to participants that gathers data to answer such research question.

\begin{RQenvironment}{1}{\RQone}
Chapters of cookbooks should have well-defined themes, which should be identifiable from the top-5 LDA terms assigned to their titles. We refer to this property as \chaptersemantics.

\begin{surveyquestionenvironment}
``Concerning the $<$five terms$>$ terms assigned to the title of the chapter $x$ from the cookbook $C$, indicate your agreement with the statement `the five terms are related with the theme of the chapter'''.
\end{surveyquestionenvironment}

\begin{responsetypeenvironment}
5-point Likert scale defined from strongly disagree (1) to strongly agree to (5):
\begin{itemize}[leftmargin=*,labelindent=1mm]
\item (5: strongly agree) It is possible to assign an unique theme to the chapter and all the five terms are related to this theme;
\item (4: agree) It is possible to assign a theme to the chapter, but some of the five terms is not related to this theme;
\item (3: neutral) Undecided or identified more than one theme;
\item (2: disagree) It it possible to see some relation between the terms, but it is not possible to assign a theme for the chapter;
\item (1: strongly disagree) It is not possible to see any type of relationship between the terms.
\end{itemize}
\end{responsetypeenvironment}
\end{RQenvironment}

\begin{RQenvironment}{2}{\RQtwo}
Recipes within a chapter should be related to the theme of the chapter. For instance, if a recipe is within a chapter about ``String Manipulation'', it must address this theme. We refer to this property as \reciperelatednesstochapter.

\begin{surveyquestionenvironment}
``Concerning the recipe $x$ from the cookbook $C$, indicate your agreement with the statement `the recipe is related to the terms assigned to the title of its chapter'''.
\end{surveyquestionenvironment}

\begin{responsetypeenvironment}
5-point Likert scale defined from strongly disagree (1) to strongly agree to (5):
\begin{itemize}[leftmargin=*,labelindent=1mm]
\item (5: strongly agree) The recipe is related to three or more terms;
\item (4: agree) The recipe is related to two terms;
\item (3: neutral) Undecided;
\item (2: disagree) The recipe is related to one term;
\item (1: strongly disagree) The recipe is not related with any term.
\end{itemize}
\end{responsetypeenvironment}
\end{RQenvironment}

\begin{RQenvironment}{3}{\RQthree}
Recipes are suitable to be part of a cookbook if they contain information from \textit{how-to-do-it} questions. Therefore, recipes where the question is of another type, such as where the asker asks for help to fix a bug, are not suitable for being in a cookbook. We refer to this property as \recipeadequacy.

 \begin{surveyquestionenvironment}
``Concerning the recipe $x$ from the cookbook $C$, indicate your agreement with the statement `the recipe is adequate to be part of the cookbook as it has \textit{how-to-do-it} nature'''.
\end{surveyquestionenvironment}

\begin{responsetypeenvironment}
5-point Likert scale defined from strongly disagree (1) to strongly agree to (5):
\begin{itemize}[leftmargin=*,labelindent=1mm]
\item (5: strongly agree) The recipe is strongly adequate;
\item (4: agree) The recipe is partially adequate;
\item (3: neutral) Undecided;
\item (2: disagree) The recipe is inadequate;
\item (1: strongly disagree) The recipe is strongly inadequate.
\end{itemize}
\end{responsetypeenvironment}
\end{RQenvironment}

\begin{RQenvironment}{4}{\RQfour}
Recipes should be self-contained, i.e., the information inside them should be sufficient to understand the described problems and their solutions. Many recipes have links to external sources (e.g., the official API documentation website, blogs, forums), and although this type of content may be beneficial as they serve as a complement to the presented solution, it is not desirable that the solution is available only in an external resource, because there is no guarantee that it will be still available in the future. We refer to this property as \recipeselfcontainment.

\begin{surveyquestionenvironment}
``Concerning the recipe $x$ from the cookbook $C$, indicate your agreement with the statement `the recipe is self-contained as it contains the necessary information to understand the scenario and its solution'''.
\end{surveyquestionenvironment}

\begin{responsetypeenvironment}
5-point Likert scale defined from strongly disagree (1) to strongly agree to (5):
\begin{itemize}[leftmargin=*,labelindent=1mm]
\item (5: strongly agree) It is possible to completely understand the scenario and its solution contained in the recipe by just considering the textual content of the recipe (disregarding the possible external sources referenced in the recipe);
\item (4: agree) It is possible to understand the most part of the scenario and its solution contained in the recipe by just considering the textual content of the recipe (disregarding the possible external sources referenced in the recipe);
\item (3: neutral) Undecided;
\item (2: disagree) To understand the most part of the scenario and its solution contained in the recipe, it is necessary to access the external sources referenced in the recipe;
\item (1: strongly disagree) The information contained in the recipe is completely dependent of the external sources referenced in the recipe.
\end{itemize}
\end{responsetypeenvironment}
\end{RQenvironment}

\begin{RQenvironment}{5}{\RQfive}
Ideally, recipes should meet more than one property as each one corresponds to a different advantage. If a recipe does not meet one of these properties, it probably should not be considered to be part of the cookbook, even if the other two properties are well-evaluated.
\end{RQenvironment}

\begin{RQenvironment}{6}{\RQsix}
This research question aims to identify the perception of the participants regarding the usefulness of crowd cookbooks in API learning.

\begin{surveyquestionenvironment}
``Would you use the cookbook to learn the API $x$? If yes, how? If no, justify your answer and, if it is possible, suggest how to enhance it''.
\end{surveyquestionenvironment}

\begin{responsetypeenvironment}
Free natural language text.
\end{responsetypeenvironment}
\end{RQenvironment}

\begin{RQenvironment}{7}{\RQseven}
This research question aims to identify the perception of the participants regarding the strengths, weaknesses and opportunities for improvement of the crowd cookbooks.

\begin{surveyquestionenvironment}
``In your opinion, which are the positive and negative points concerning the crowd cookbooks? Also, suggest improvements for the generation of the cookbooks''.
\end{surveyquestionenvironment}

\begin{responsetypeenvironment}
Free natural language text.
\end{responsetypeenvironment}
\end{RQenvironment}

\subsection{API subjects}\label{sec:API-subjects}

We generated cookbooks for three APIs: SWT\footnote{http://www.eclipse.org/swt/} (Java), LINQ\footnote{http://msdn.microsoft.com/pt-br/library/bb397926.aspx} (.NET programming languages) and QT\footnote{http://qt-project.org/} (C++). These APIs are widely used by the software development community, and they are related to different programming languages, which allows us to observe, to some extent, the generality of our approach.

The size of the generated cookbooks are: 12 chapters and 69 recipes for SWT; 9 chapters and 94 recipes for LINQ; and 12 chapters and 119 recipes for QT.
The evaluation of all the chapters and recipes from the cookbooks would take a long time due to their sizes. For this reason, participants evaluated a stratified sample of chapters and recipes from the cookbooks---the sampling process is presented in Section~\ref{sec:Sampling-chapters-and-recipes-for-evaluation}.

\subsection{Human subjects}\label{sec:Human-subjects}

The human subjects that participated in the study to assess the cookbooks were graduate students in Computer Science (MSc or PhD) from the Federal University of Uberl\^andia and/or junior software development professionals from Uberl\^andia city. All of them work with software development and have similar profile in terms of experience.

Initially, 33 people expressed interest in participating in the study. Aiming to prepare the participants for the evaluation process, we prepared a training document containing information on the purpose of the study and instructions on how to proceed during the assessment of cookbooks. 
Out of the 33 initial volunteers, 16 completed the evaluation process, which consisted of filling out questionnaires built using the online service LimeSurvey\footnote{http://www.limesurvey.org/en}.

To check the expertise level of the participants on the APIs considered in the study, we asked them to inform their respective level for each API (novice, intermediary or advanced), if any. Most of the subjects had no knowledge on the APIs. Exceptions were that out of 16 subjects, four answered {\it novice} for SWT, one answered {\it intermediate} for LINQ, and one answered {\it novice} for QT.

\subsection{Construction of controlled cookbooks}\label{sec:Construction-of-controlled-cookbooks}

To ensure that participants were being honest during the evaluation and that they had understood the evaluated properties, we built modified versions of the cookbooks. These modified versions are called \textit{controlled cookbooks}.

So, the set of cookbooks considered in this study is: SWT-Cookbook, SWT-Controlled-Cookbook, LINQ-Cookbook, LINQ-Controlled-Cookbook, QT-Cookbook and QT-Controlled-Cookbook. The non-controlled cookbooks were constructed with Algorithm 1. The controlled cookbooks were manually constructed from non-controlled cookbooks adding items (i.e., recipes and chapters) that are notoriously bad to the original cookbooks, with respect to the defined properties. For the \chaptersemantics~property, we created a chapter title with five unrelated terms. For instance, in the case of SWT-Controlled-Cookbook, we created a chapter title: ``pdf press thread eclipse wizard''. For each other property (i.e., \reciperelatednesstochapter, \recipeadequacy~and~\recipeselfcontainment), we have chosen a Q\&A pair that is bad regarding these properties. For example, for the \recipeadequacy~property, we picked a Q\&A in which the question is not a \textit{how-to-do-it}. Since each item artificially included in the controlled cookbooks is bad with respect to one property, ideally they should be rated with disagreement values (rating values 1 or 2).

\subsection{Sampling chapters and recipes for evaluation}\label{sec:Sampling-chapters-and-recipes-for-evaluation}

The assessment of all chapters and recipes from all cookbooks for all research questions would be a very time-demanding task. For this reason, we conducted a stratified sampling of chapters and recipes.
Given a cookbook $C$ for the API $x$, both recipe and chapter samplings are conducted simultaneously for the non-controlled cookbook and for the controlled cookbook. Recipe sampling is performed as follows:

\begin{enumerate}
\item For both cookbooks, randomly sample one recipe from each chapter of the \textit{cookbook} $C$, resulting $n$ sampled recipes. Thus, the random sample is stratified by chapter, guaranteeing that a representative part of the cookbook is sampled.

\item For the controlled cookbook, additionally to those $n$ sampled recipes, include four \textit{control} recipes. Recall that each one of these control recipes is a specifically bad recipe according to one of the four properties related to recipes mentioned above.

\item For the non-controlled cookbook, additionally to those $n$ sampled recipes, sample four other recipes with no other criteria, so that both controlled and non-controlled sampled cookbooks have $n+4$ recipes.
\end{enumerate}

Chapter sampling is performed as follows:

\begin{enumerate}
\item Chapters are sorted by number of recipes.

\item For both cookbooks, randomly sample two chapters with number of recipes lower than the median number of recipes and two chapters with number of recipes higher than the median. After this step, four chapters have been sampled.

\item For the controlled cookbook, include one controlled chapter, so the controlled cookbook has five chapters and the non-controlled cookbook has four chapters.
\end{enumerate}

\subsection{Participant assignment to control cookbooks}\label{sec:Participant-assignment-to-control-cookbooks}

All study participants evaluated the sampled cookbooks showed in the previous subsection answering to Likert-scale questions and essay questions. However, since we created cookbooks with controlled items, this could influence answers to the essay questions: evaluating cookbooks with artificially included bad items would not make sense. For this reason, the participants were divided into three groups: SWT-Controlled, LINQ-Controlled and QT-Controlled. Note that participants did not know about the existence of controlled items.

Participants in SWT-Controlled rated the cookbooks SWT-Controlled-Cookbook, LINQ-Cookbook and QT-Cookbook, but the essay questions were just about the two non-controlled cookbooks (LINQ-Cookbook and QT-Cookbook). Similar logic was applied to the other two groups. The participants were randomly assigned to one of the three groups in a way that the sizes of the groups are balanced. The groups SWT-Controlled and QT-Controlled had 5 subjects each, and LINQ-Controlled had 6 subjects.

\subsection{Answers to controlled items}\label{sec:Responses-to-controlled-items}

The answers given to questions related to controlled items are shown in Table~\ref{table:responses-controlled-items}, grouped by API. In this table, ratings with disagreement values were colored in different shades of gray: dark gray for 1 (strongly disagree) and light gray for 2 (disagree). Ideally, all the cells from such table should be gray, as the controlled chapters and the controlled recipes are not good. In fact, we noted that most of the cells are gray, which means that the controlled items were mostly assessed with disagreement values by the participants.

\begin{table}[ht]
  \caption{Responses to the controlled items of the three APIs.}
  \label{table:responses-controlled-items}
  \centering
  \small
  \begin{threeparttable}
    \begin{tabular}{x{.035\textwidth} p{.039\textwidth} x{.065\textwidth} x{.065\textwidth} x{.065\textwidth} x{.07\textwidth} }
      \toprule
      Sub- & \multirow{2}{*}{API} & \textit{Chapter} & \textit{Recipe} & \textit{Recipe} & \textit{Recipe} \\
      ject & & \textit{Semantics} & \textit{Adequacy} & \textit{SelfCont} & \textit{Relatedness} \\
      \midrule
      4 & SWT & 1\cellcolor{dark-red} & 2\cellcolor{medium-red} & 1\cellcolor{dark-red} & 1\cellcolor{dark-red} \\
      5 & SWT & 1\cellcolor{dark-red} & 1\cellcolor{dark-red} & 1\cellcolor{dark-red} & 1\cellcolor{dark-red} \\
      8 & SWT & 2\cellcolor{medium-red} & 1\cellcolor{dark-red} & 1\cellcolor{dark-red} & 1\cellcolor{dark-red} \\
      10 & SWT & 1\cellcolor{dark-red} & 1\cellcolor{dark-red} & 1\cellcolor{dark-red} & 1\cellcolor{dark-red} \\
      16 & SWT & 1\cellcolor{dark-red} & 1\cellcolor{dark-red} & 1\cellcolor{dark-red} & 2\cellcolor{medium-red} \\
      3 & LINQ & 3 & 5\cellcolor{dark-green} & 2\cellcolor{medium-red} & 4\cellcolor{medium-green} \\
      9 & LINQ & 1\cellcolor{dark-red} & 2\cellcolor{medium-red} & 3 & 3 \\
      11 & LINQ & 1\cellcolor{dark-red} & 1\cellcolor{dark-red} & 1\cellcolor{dark-red} & 1\cellcolor{dark-red} \\
      12 & LINQ & 3 & 3 & 1\cellcolor{dark-red} & 3 \\
      14 & LINQ & 2\cellcolor{medium-red} & 1\cellcolor{dark-red} & 1\cellcolor{dark-red} & 2\cellcolor{medium-red} \\
      15 & LINQ & 1\cellcolor{dark-red} & 1\cellcolor{dark-red} & 1\cellcolor{dark-red} & 1\cellcolor{dark-red} \\
      1 & QT & 3 & 3 & 3 & 1\cellcolor{dark-red} \\
      2 & QT & 3 & 4\cellcolor{medium-green} & 3 & 4\cellcolor{medium-green} \\
      6 & QT & 2\cellcolor{medium-red} & 1\cellcolor{dark-red} & 2\cellcolor{medium-red} & 2\cellcolor{medium-red} \\
      7 & QT & 2\cellcolor{medium-red} & 4\cellcolor{medium-green} & 1\cellcolor{dark-red} & 2\cellcolor{medium-red} \\
      13 & QT & 1\cellcolor{dark-red} & 2\cellcolor{medium-red} & 2\cellcolor{medium-red} & 2\cellcolor{medium-red} \\
      \bottomrule
    \end{tabular}
    \begin{tablenotes}
      \scriptsize
      \item \fivepointLikertscale.
    \end{tablenotes}
  \end{threeparttable}
\end{table}

The participants who did rate the controlled items with agreement values (4 or 5) or neutral (3) for a given property (e.g. \chaptersemantics) were excluded in the analysis of the results presented in this paper based on the non-controlled items (Section \ref{sec:results-and-discussion}). For example, all the ratings of subject 4 are considered; however, only the ratings for \chaptersemantics~and~\recipeadequacy~properties by subject 9 are considered; and none of the ratings given by subject 2 is considered.

\section{Evaluation Results and Discussion}\label{sec:results-and-discussion}

In this section, we present and discuss the results for each research question. The results are on the answers given by the participants to non-controlled items. Moreover, the results for each property (e.g. \chaptersemantics) are on the answers from participants that negatively evaluated such property in the analysis of controlled items as described in Section \ref{sec:Responses-to-controlled-items}.

\subsection{On the \chaptersemantics~property (RQ \#1)}

\begin{contextenvironment}
12 participants rated 12 chapters each, according to the 5-point Likert scale for the \chaptersemantics~property, resulting 144 ratings.
\end{contextenvironment}

\begin{findingsenvironment}
The ratings are distributed in the 5-point Likert scale as follows: 10.42\% is strongly disagree, 14.58\% is disagree, 15.28\% is neutral, 24.31\% is agree, and 35.42\% is strongly agree. Thus, 59.72\% of the 144 ratings are from participants who agreed that the five terms of a given chapter are related with the theme of the chapter as they could derive this theme through these terms.
\end{findingsenvironment}

\begin{statisticsenvironment}
Figure \ref{fig:violinplot-chapter_semantics} shows, for each of the five possible rating values for the \chaptersemantics~property, the distribution of the number of ratings.  Kruskal-Wallis test on the rating values returned \textit{p-value} = 0.011.
Post-hoc analysis with pairwise comparisons using Tukey and Kramer (Nemenyi) test showed that 
the number of rating value 1 is significantly lower than the number of rating value 5.

\begin{figure}[h!]
  \centering
  \includegraphics[scale=0.58]{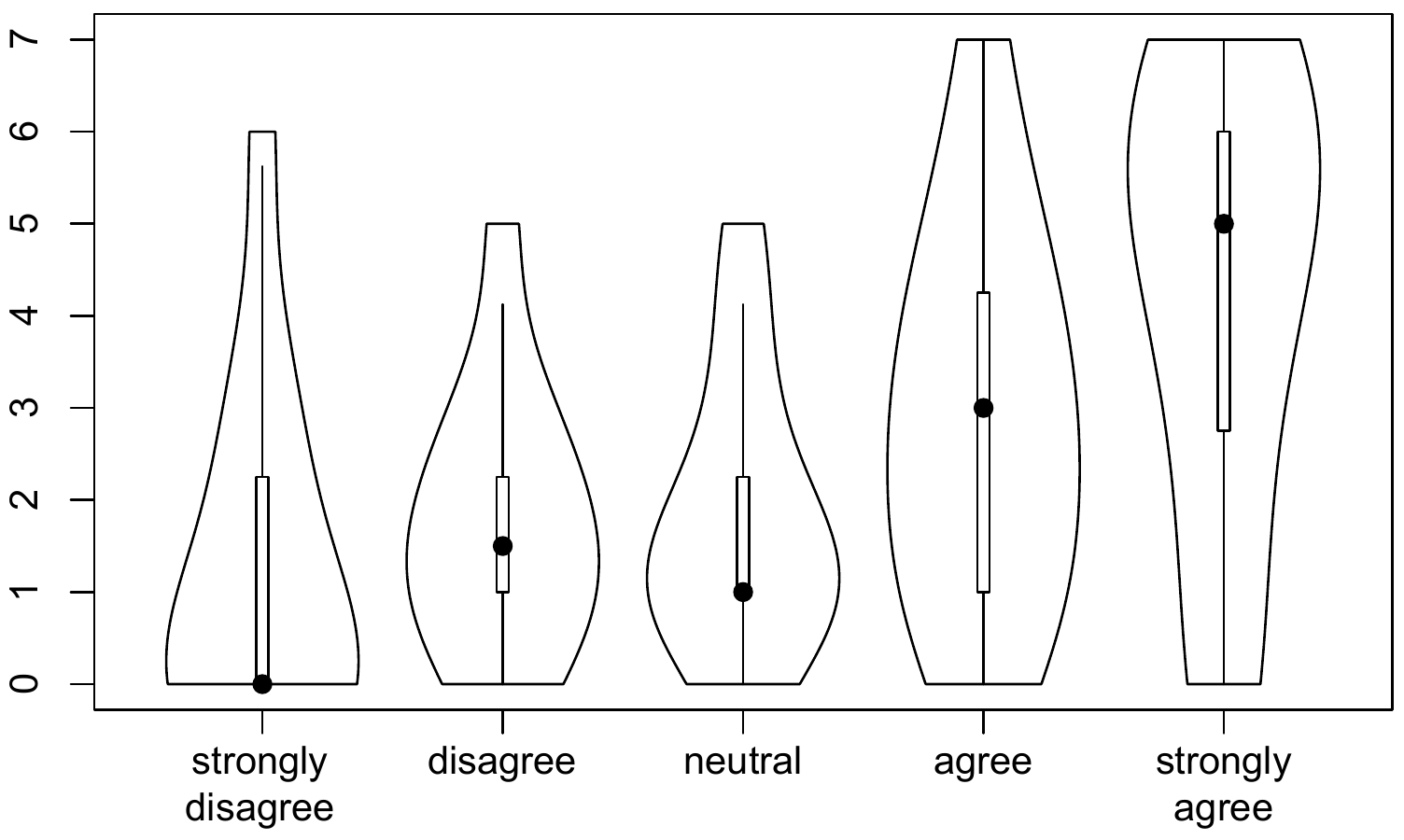}
  \caption{Response frequency distribution for the \chaptersemantics~property per rating.}
  \label{fig:violinplot-chapter_semantics}
\end{figure}

\end{statisticsenvironment}

\begin{participantscommentsenvironment}
From comments made by participants, one of the opportunities for improvement is to form the chapter titles with the original words instead of stems (i.e., ``table'' instead of ``tabl'').
\end{participantscommentsenvironment}

\begin{mdframed}[style=MyFrame]
\RQcitation{1}\textbf{:} \textit{\RQone} \\
59.72\% of the ratings are from participants that could derive a theme for chapters through the five terms assigned to the chapters, i.e., they agreed that the chapters have a defined theme.
\end{mdframed}

\subsection{On the \reciperelatednesstochapter~property (RQ \#2)}

\begin{contextenvironment}
12 participants rated 41 recipes each, according to the 5-point Likert scale for the \reciperelatednesstochapter~property, resulting 492 ratings.
\end{contextenvironment}

\begin{findingsenvironment}
The ratings are distributed in the 5-point Likert scale as follows: 9.15\% is strongly disagree, 6.5\% is disagree, 8.74\% is neutral, 26.63\% is agree, and 48.98\% is strongly agree. Thus, 75.61\% of the 492 ratings are from participants who agreed that the recipe is related to the title of its chapter, i.e., that the recipe is related to at least two terms assigned to the title.
\end{findingsenvironment}

\begin{statisticsenvironment}
Figure \ref{fig:violinplot-recipe_relatedness_to_chapter} shows, for each of the five possible rating values for the \reciperelatednesstochapter~property, the distribution of the number of ratings.  Kruskal-Wallis   test returned \textit{p-value} = $6.8 \times 10^{-7}$. 
Post-hoc analysis with pairwise comparisons using Nemenyi test
showed 
that the number of rating value 5 is significantly higher than the number of rating values 1 and 2. There was no significant difference from rating value 5 to rating value 4, but still the number of rating value 4 is significantly higher than rating value 2, but not with rating value 1.

\begin{figure}[h!]
  \centering
  \includegraphics[scale=0.58]{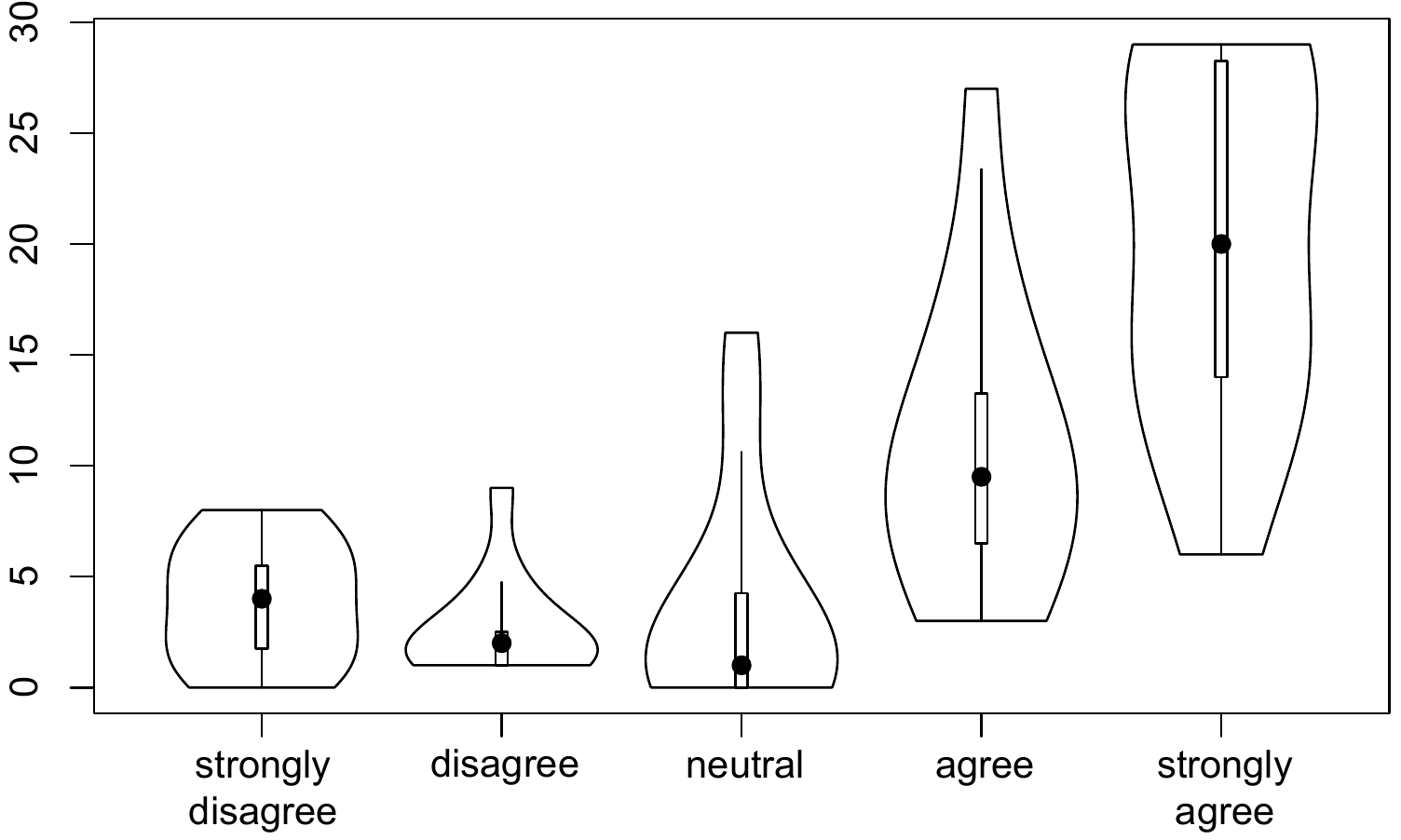}
  \caption{Response frequency distribution for the \reciperelatednesstochapter~property per rating.}
  \label{fig:violinplot-recipe_relatedness_to_chapter}
\end{figure}

\end{statisticsenvironment}

\begin{participantscommentsenvironment}
From comments made by participants, the main criticism regarding the \reciperelatednesstochapter~property is the existence of a few poorly localized recipes, i.e., recipes that are in certain chapters, but that are more related to the theme of other chapters.
\end{participantscommentsenvironment}

\begin{additionalanalysisenvironment}
Concerning the \reciperelatednesstochapter~property, we provide additional data on how chapters overlap. We calculated, for each chapter, the number of terms of its title that are also part of the title of other chapters. 
The presence of repeated terms between chapters could be an indication of the existence of the same theme being treated in more than one chapter, in a larger or lower extent.

We found that the number of chapters having terms in common with other chapters is not high. In the case of SWT, there are only six pairs of chapters with non-empty intersection, corresponding to 9\% of all possible pairs of chapter (66 pairs, i.e., 2-combination of 12 chapters). In the case of LINQ, we had 16.67\%, and 9\% in the case of QT.  

This additional data analysis also corroborates with the survey result, because the low intersection in terms from chapters' titles indicates a low number of posts on the intersected themes, therefore reducing chances of recipe non-relatedness to chapter.
\end{additionalanalysisenvironment}

\begin{mdframed}[style=MyFrame]
\RQcitation{2}\textbf{:} \textit{\RQtwo} \\
75.61\% of the ratings are from participants that agreed that the recipe is related to the theme of its chapter.
\end{mdframed}

\subsection{On the \recipeadequacy~property (RQ \#3)}

\begin{contextenvironment}
11 participants rated 41 recipes each, according to the 5-point Likert scale for the \recipeadequacy~property, resulting 451 ratings.
\end{contextenvironment}

\begin{findingsenvironment}
The ratings are distributed in the 5-point Likert scale as follows: 4.43\% is strongly disagree, 4.66\% is disagree, 7.54\% is neutral, 27.72\% is agree, and 55.65\% is strongly agree. Thus, 83.37\% of the 451 ratings are from participants who agreed that the recipe is adequate to be part of the cookbook.
\end{findingsenvironment}

\begin{statisticsenvironment}
Figure \ref{fig:violinplot-recipe_adequacy} shows, for each of the five possible rating values for the \recipeadequacy~property, the distribution of the number of ratings.  Kruskal-Wallis  test returned \textit{p-value} = $4.3 \times 10^{-6}$. 
The result of the post-hoc analysis with pairwise comparisons using Nemenyi test 
showed 
that the number of rating value 5 is significantly higher than the number of rating values 1, 2 and 3. There was no significant difference from rating value 5 to rating value 4, but still the number of rating value 4 is significantly higher than rating value 1, but not with rating value 2.

\begin{figure}[h!]
  \centering
  \includegraphics[scale=0.58]{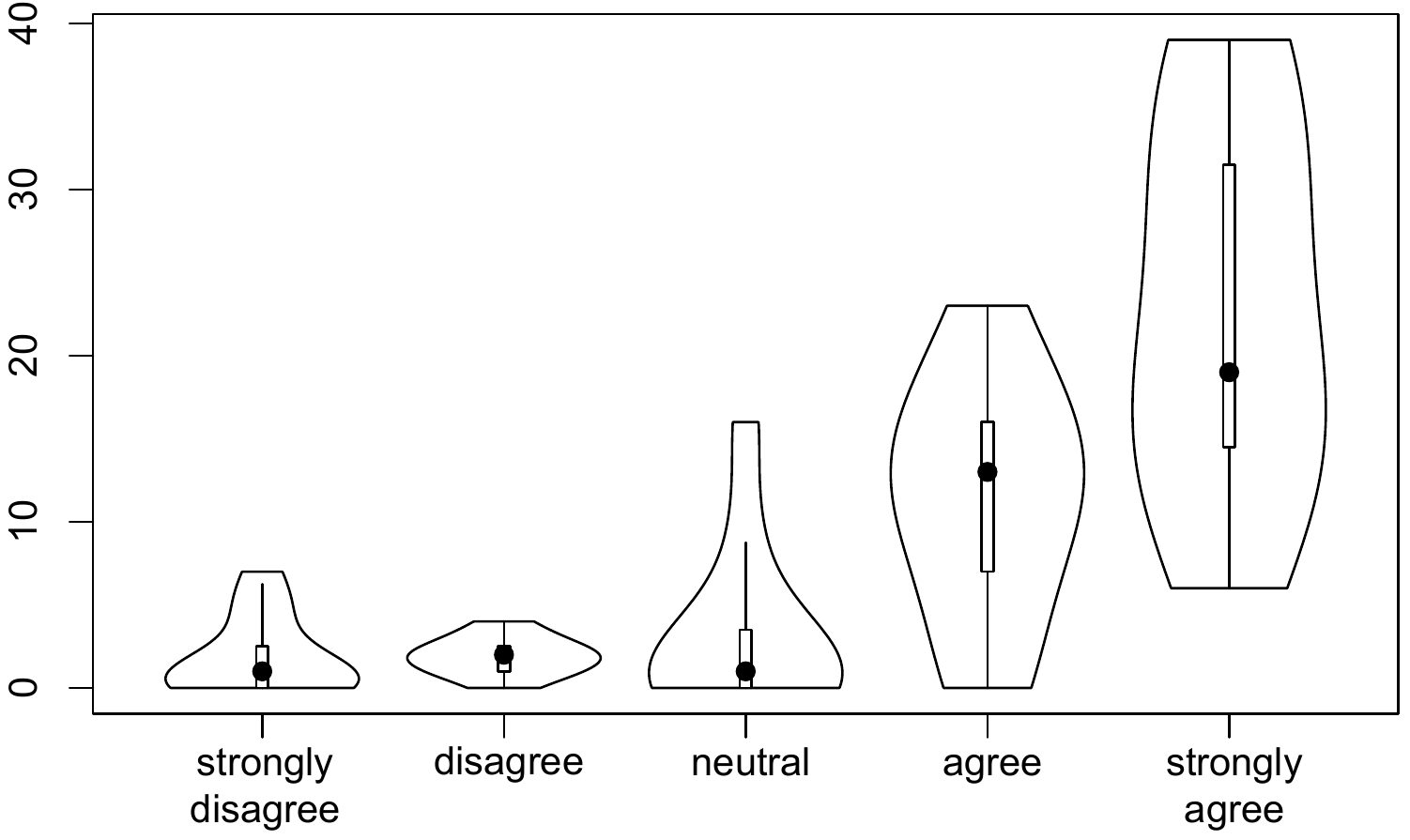}
  \caption{Response frequency distribution for the \recipeadequacy~property per rating.}
  \label{fig:violinplot-recipe_adequacy}
\end{figure}

\end{statisticsenvironment}

\begin{participantscommentsenvironment}
The participants pointed out several factors that make a given recipe suitable for cookbooks: the presence of explanations of the API elements used in code snippets; questions that are \textit{how-to-do-it}; the presence of references to external sources for additional information; and the presence of explanations in natural language.
\end{participantscommentsenvironment}

\begin{mdframed}[style=MyFrame]
\RQcitation{3}\textbf{:} \textit{\RQthree} \\
83.37\% of the ratings are from participants that agreed that the recipe is adequate to be part of the cookbook.
\end{mdframed}

\subsection{On the \recipeselfcontainment~property (RQ \#4)}

\begin{contextenvironment}
13 participants rated 41 recipes each, according to the 5-point Likert scale for the \recipeselfcontainment~property, resulting 533 ratings.
\end{contextenvironment}

\begin{findingsenvironment}
The ratings are distributed in the 5-point Likert scale as follows: 0.94\% is strongly disagree, 6\% is disagree, 5.25\% is neutral, 18.76\% is agree, and 69.04\% is strongly agree. Thus, 87.8\% of the 533 ratings are from participants who agreed that the recipe is self-contained as external sources referenced in the recipe are not required to understand the scenario and its solution presented in the recipe.
\end{findingsenvironment}

\begin{statisticsenvironment}
Figure \ref{fig:violinplot-recipe_self_containment} shows, for each of the five possible rating values for the \recipeselfcontainment~property, the distribution of the number of ratings.  Kruskal-Wallis  test returned \textit{p-value} = $1.804 \times 10^{-9}$.
Post-hoc analysis with pairwise comparisons using Nemenyi test 
showed that the number of rating value 5 is significantly higher than the number of rating values 1, 2 and 3. There was no significant difference from rating value 5 to rating value 4, but still the number of rating value 4 is significantly higher than rating value 1, but not with rating value 2.

\begin{figure}[h!]
  \centering
  \includegraphics[scale=0.58]{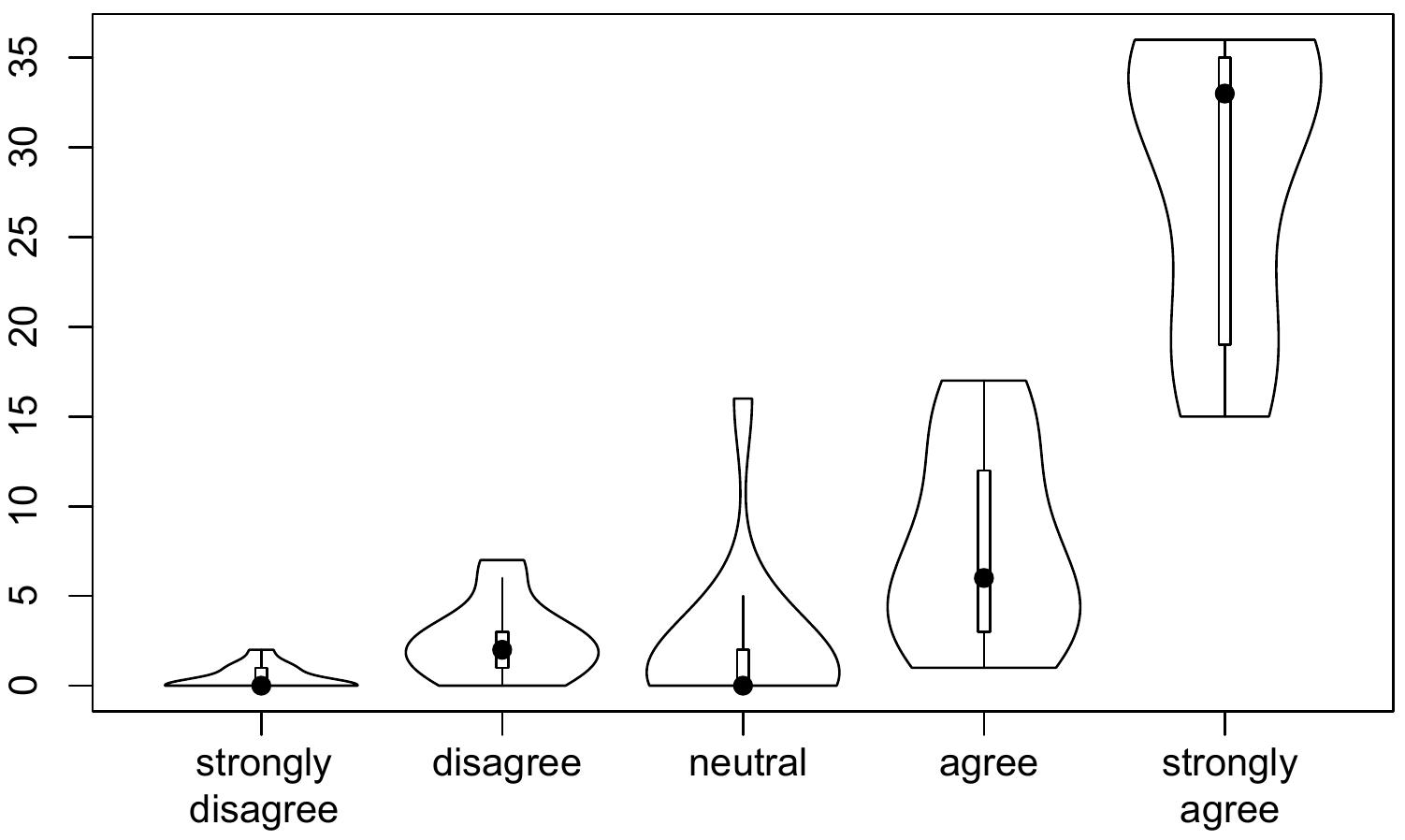}
  \caption{Response frequency distribution for the \recipeselfcontainment~property per rating.}
  \label{fig:violinplot-recipe_self_containment}
\end{figure}

\end{statisticsenvironment}

\begin{participantscommentsenvironment}
From the comments made by participants, we concluded that the presence of very summarized recipes makes the comprehension of the information contained in them difficult, and the participants claimed the need of additional sources of information in such cases.
\end{participantscommentsenvironment}

\begin{mdframed}[style=MyFrame]
\RQcitation{4}\textbf{:} \textit{\RQfour} \\
87.8\% of the ratings are from participants that agreed that the recipe is self-contained as it contains the necessary information to understand the scenario and its solution, i.e., external sources referenced in the recipe are not required.
\end{mdframed}

\subsection{On the join of \reciperelatednesstochapter, \recipeadequacy~and~\recipeselfcontainment~properties (RQ \#5)}

Recipes should have good evaluation with respect to more than one property simultaneously. Thus, for those participants that had the \reciperelatednesstochapter, \recipeadequacy~and~\recipeselfcontainment~columns painted gray in Table \ref{table:responses-controlled-items}, we checked the number of recipes that were rated with values 4 or 5 for all the three properties related to recipes. The \chaptersemantics~property was not considered in this case since it applies to chapters instead of recipes. The average number of recipes rated with value 4 or 5 for the three properties is 26.8, which corresponds to 65.37\% of assessed recipes for each subject.

\begin{mdframed}[style=MyFrame]
\RQcitation{5}\textbf{:} \textit{\RQfive} \\
The percentage of recipes rated with values 4 and 5 for the three properties related to recipes (\reciperelatednesstochapter, \recipeadequacy~and~\recipeselfcontainment) is 65.37\%, which is high.
\end{mdframed}

\subsection{On the usage of cookbooks for API learning (RQ \#6)}

\begin{contextenvironment}
Participants evaluated cookbooks with free natural language text.
\end{contextenvironment}

\begin{findingsenvironment}
For SWT API, answers were collected from ten subjects. Four of them said they would use the SWT-Cookbook to learn about such API. Six subjects said they would prefer to use the cookbook just to seek information about specific problems. In this case, the cookbooks would not be used as an introductory documentation to learn an API as a whole, from the beginning to the end. The lack of a logical sequence between chapters and the lack of a ``Hello-Word'' recipe were cited as something that hinders the usage of cookbooks to learn an API. The fact that the cookbook has several basic information about the API and the presence of source code that can be executed were cited as justifications to use the cookbook for learning about SWT.

For LINQ API, answers were collected from nine subjects. Six of them said that the cookbook could be used to solve specific problems, although three of them have said that the cookbook could also be used in learning. The organization of the cookbook in chapters was cited as an intuitive factor that facilitates learning about the API, since Stack Overflow originally does not have this structure. The presence of generic questions was also cited as a positive aspect of using the cookbook for learning.

For QT, answers were collected from eleven subjects. We also observed that a prevalent opinion (five out of eleven) is that the cookbook could be used to obtain information about specific problems but not to learn about the API from the beginning to the end.

Overall, the prevalent view from the answers from the participants on the three cookbooks is that they can be used as a source of information for specific problems but not for learning an API as a whole. The main reason cited by the participants on not using cookbooks for learning an API as a whole is the lack of logical and chronological order among cookbook chapters.

From all of the 30 answers, 17 were from participants who said they would use cookbooks to seek information about a specific problem. Indeed, it is possible for  developers  navigating through the chapters and recipes of the cookbooks to locate needed information, since the cookbooks have only two levels of hierarchy (chapters and recipes), which facilitates their exploration. A possible usage would be checking if the cookbook has a chapter theme connected with the problem theme, and if there is such a chapter, peruse the recipes contained therein. However, it is possible that there is no recipe in the cookbook that is specific to the current problem, because the cookbooks are built from content well evaluated by the crowd on Stack Overflow, and do not target high API coverage. 

Additionally, there could be more efficient ways to seek information about a specific problem as, for example, searching with Google, the Stack Overflow search engine, or even using recommendation approaches that use Stack Overflow data~\cite{Ponzanelli2013,Souza2014,Campos2016}. We advocate that due to its \textit{browsing} nature, cookbooks can be complementary to \textit{searching} mechanisms. Browsing can be used in cases where it is not possible to formulate keywords for composing a query, which can occur, for example, when a developer is relatively inexperienced concerning the API~\cite{Olston2003}.
\end{findingsenvironment}

\begin{mdframed}[style=MyFrame]
\RQcitation{6}\textbf{:} \textit{\RQsix} \\
The most prevalent statement among the participants (17 out of 30 answers) is that crowd cookbooks could be used to seek information about a specific problem. Since the cookbook construction approach has the ability to filter, summarize and organize recipes, we conclude that crowd cookbooks could also be indicated to help new developers to become familiarized with the API features and vocabulary, to be able to produce queries when trying to solve specific problems.
\end{mdframed}

\subsection{On strengths, weaknesses and suggested improvements (RQ \#7)}

\begin{contextenvironment}
Participants evaluated cookbooks with free natural language text.
\end{contextenvironment}

\begin{findingsenvironment}
The first author of this paper conducted a \emph{Thematic Analysis}~\cite{Cruzes2011} to summarize the answers given by the participants with respect to the strengths, weaknesses and improvements that can be made to the generated cookbooks. Such analysis comprehends reading all answers, identifying specific segments of text, coding the segments, reducing overlaps, and mapping codes into the following themes which were considered sufficiently synthesized.

The main strengths cited were:
1) presence of chapters with well-defined themes;
2) organization that facilitates locating important information, since Stack Overflow originally lacks the organization available in cookbooks;
3) presence of answers that refer to additional sources of information;
4) presence of generic problems;
and 5) existence of source code examples in recipes.

The main weaknesses cited were:
1) presence of very specific questions;
2) presence of recipes that are not related to the theme of their chapters;
3) presence of chapters with poorly-defined themes;
4) presence of stems in the titles of the chapters instead of original words;
5) presence of chapters with too many recipes;
and 6) presence of very brief/incomplete recipes and with too small/incomplete source code.

The improvements suggested were:
\begin{itemize}
\item The recipes could contain the indications for other similar recipes. Although each chapter brings together recipes on the same general theme, it would be interesting to know which other recipes deal with similar problems, at a lower level of granularity;
\item The recipes could explain the used classes/methods, so less experienced developers would not need to use other sources for this information. A way to achieve such feature could be to identify the API elements (e.g., classes and methods) used in the source code snippets in the recipe, and to link them to the official API documentation. For example, we could use the approach presented by Subramanian et al.~\cite{Subramanian2014};
\item The cookbooks could have a chronological sequence to present chapters, for example, from basic chapters to more advanced ones. However, chronological sequence of chapters is not a typical characteristic of cookbooks. This feature is more common in other kinds of documentation, such as tutorials.
\end{itemize}
\end{findingsenvironment}

\begin{mdframed}[style=MyFrame]
\RQcitation{7}\textbf{:} \textit{\RQseven} \\
Among the points raised by the participants, the highlights are: as positive point, that the cookbooks have an easy organization for locating important information, since Stack Overflow originally lacks that organization; as negative point, the presence of recipes that are not related to the theme of their chapters, and the presence of chapters with poorly-defined themes; and as suggestion for improvements, to indicate in recipes, other similar recipes, and to include information on the API elements used in a recipe.
\end{mdframed}

\section{Limitations and Threats to Validity}\label{sec:threats}

There are some limitations in our approach to construct cookbooks.
First, although the approach relies on standard LDA technique with long history in software engineering academic research~\cite{Chen2016emse}, we use LDA without sophisticated parameter tuning to generate topics, which might have a detrimental effect on the generated topics.
Recent study~\cite{Agrawal2018} suggests that the order of inputed texts to train LDA matters, i.e., topics may vary given different order of the same input.

Another limitation is that the complexity of chapters and recipes was not taken into account, so chapters and recipes are currently not organized according to the complexity of understanding. 
How to rank chapters and recipes according to their complexities could be addressed in future work.

There are some threats to internal validity of the results. The participants of the study are mostly acquaintances of the first author of this paper, which could create a bias towards evaluation, because they could positively evaluate the cookbooks just trying to benefit the research results. To mitigate this threat, we created controlled cookbooks that contain chapters and recipes notoriously bad, and we checked how the subjects rated these items (Section \ref{sec:Responses-to-controlled-items}).

Participants that evaluated the cookbooks were typically not familiar with the APIs, which would be a severe threat if they had to evaluate the correctness of the content in recipes. However, the recipes contain content from Stack Overflow, and our approach relies on the score of the posts on Stack Overflow to produce a reliable cookbook. Moreover, the participants were not evaluating the correctness of the posts. The evaluation point of view is from a person that does not know the technical details of the API, but could be capable to recognize whether the provided content would be relevant, organized, and coherent, so someone else could use it to use that unknown API.

There are also some threats to external validity. The study we conducted to evaluate our approach was performed with only three APIs, which may impact the generalization of the results. However, we chose APIs related to different programming languages (SWT--Java, LINQ--.NET languages and QT--C++), which minimizes this threat.

Another factor that affects the generalization of the results is that only 16 subjects participated in the evaluation, which may not correspond to a representative sample of the software development community. Moreover, the participants assessed a sample of recipes and chapters from cookbooks, instead the entire cookbooks. To minimize this threat, we randomly sampled recipes from each chapter of the cookbooks to create a representative sample of the cookbooks.

\section{Related Work}\label{sec:related-work}

Documenting and using APIs effectively is not trivial~\cite{Parnin2012}. There are some tools designed to produce or augment documentation for APIs.
Stylos et al.~\cite{Stylos2009} developed Jadeite, which extracts common usage scenarios and inserts them into an existing API documentation (Javadoc). However, the scenario examples are limited to instantiation of classes.
Kim et al.~\cite{Kim2009} proposed a technique to automatically increment the documentation of APIs (e.g., Javadocs) with source code examples. According to their results, it was possible to include code examples for more than 20,000 API methods  (around 75\% of all elements considered in the study).
Montandon et al.~\cite{Montandon2013} developed APIMiner, a platform that enhances the documentation of APIs for Java (i.e., Javadocs) with concrete usage examples, extracted from a private code repository.
There are three main differences between those works and ours: 1) our approach produces cookbooks as type of documentation, those other approaches focused on Javadocs; 2) our approach constructs documentation from scratch instead of improving existing documentation; and 3) whereas the data source for those tools are repositories of software projects, our approach uses the crowd knowledge available on Stack Overflow.

Hen$\beta$ et al.~\cite{Henbeta2012} presented a semi-automated technique to build FAQs (Frequently Asked Questions) from data available in discussions, such as mailing lists and forums. There are similarities between their approach and ours, regarding the method for producing the documentation. For instance, both strategies apply similar data preprocessing before applying LDA. However, there are differences between cookbooks and FAQs. First, cookbooks are designed to contain practical problems that developers may encounter, and we included only \textit{how-to-do-it} questions in cookbooks. In contrast, FAQs include different types of question. Second, the content available on Stack Overflow is semi-structured in questions and answers while the content in mailing listings and forums is not structured. Third, the recipes of the cookbooks must contain source code examples, since the goal of the recipes is to show how to solve programming tasks using an API. On the other hand, the goal of FAQs is to organize knowledge scattered in natural language text.

The content available on Stack Overflow has also been used in previous works to support developers at using APIs. We find two different veins of work: API documentation, which is the focus of this work, and recommendation systems. For the former, Treude and Robillard \cite{Treude2016} presented an automatic approach to mine insight sentences from Stack Overflow and augment existing API documentation (Javadoc) with them. These insight sentences are related to a particular API type (e.g. class) and provide insight not contained in the API documentation of that type. The main differences between such work and ours are: 1) our approach constructs documentation from scratch instead of improving existing documentation; and 2) their focus is on documenting API elements individually (each mined insight sentence is about an API element), while our focus is on documenting APIs with problem-solution recipes, which may include several API elements. 

On recommendation systems using Stack Overflow content, Campos et al. \cite{Campos2016} proposed a solution to recommend Q\&A pairs to programming tasks that developers are facing.
This type of work is different from our work by conception: the recommendation system proposes solutions for developers given a query (searching process), while crowd cookbooks are generated without a query for an entire API (giving an overview of the API, to be used in a browsing process). Searching and browsing are different but complementary strategies \cite{Olston2003}. Finally, crowd cookbooks are an organization of the API documentation that is spread on Stack Overflow, while recommendation systems do not actually create a documentation, but provide suggestions for developers given a context.

\section{Conclusions}\label{sec:conclusions}

In this paper, we reported on an approach to build crowd cookbooks (recipe-oriented book) to document APIs using the crowd knowledge available on Stack Overflow. 
Our approach structures chapters using LDA and fills them out with recipes, each one composed by a question and an answer from Stack Overflow posts. We selected a special kind of question (\textit{how-to-do-it}) and well voted posts from Stack Overflow to be included in cookbooks. The proposed approach applies LDA on all how-to-do-it content of an API, which results a partitioning of the content of the corpus in a broad range of dissimilar topics.

We conducted an evaluation of our cookbook generating approach to check the quality of chapter organization and selected recipes that the approach can produce.
We generated cookbooks for three APIs (SWT, LINQ and QT), and defined desirable properties that chapters and recipes must meet. Finally, human subjects (participants) evaluated the generated cookbooks based on the defined properties.

Among the findings, we highlight that the participants found a considerable percentage of the cookbooks' chapters having well-defined themes (59.72\% positive ratings). On the recipes, the participants found that the majority of recipes are related to the terms of the chapter that they are part of (75.61\% positive ratings), are suitable to be in the cookbook (83.37\% positive ratings), and have self-contained information (87.8\% positive ratings). 

Despite the fact that cookbooks were originally designed to be used through an exploration (browsing) strategy, we concluded that they can be useful for an API newcomer to become familiar with the terms of that API, and then being able to use a searching strategy more effectively to obtain specific information about a problem in hand.

Although human-edited versions of cookbooks may typically have better quality than our crowd cookbooks, it is not for every type of technology or API that there exists commercial version of cookbooks. Thus, crowd cookbooks can be especially useful for APIs that lack official documentation and/or cookbooks produced/edited by humans. Moreover, they could be a useful starting point to manually edit polished versions of cookbooks. Moreover, after setting the environment, cookbooks can be generated straightforwardly: we provide a site\footnote{\url{http://lascam.facom.ufu.br:8080/cookbooks}} with several examples of automatically generated cookbooks with default parameters, such as, target number of chapters equals 15 and initial maximum rank allowed equals 200.

As future work, techniques for tuning LDA parameters to provide topics more adherent to chapters and recipes could be proposed.
Furthermore,  other metrics (e.g., cosine similarity) to enhance the recipe relatedness to its chapter theme could be investigated. There are other research opportunities related to software documentation, for instance, to produce other kinds of documentation (e.g., tutorials) from the \textit{crowd knowledge}.

\section*{Acknowledgment}

We acknowledge CAPES, CNPq, and FAPEMIG for partial funding. We  acknowledge the  students and professionals who participated on this study.

\section*{References}

\bibliographystyle{elsarticle-num}
\bibliography{references}

\end{document}